\documentclass{article}
\usepackage{amscd,amsmath,amssymb,bbm}
\usepackage{amsfonts}
\newcommand{\p}[1]{(\ref{#1})}

\newcommand{\bxi}{{\bar\xi}}

\newcommand{\bpsi}{{\bar\psi}}
\newcommand{\bchi}{{\bar\chi}}
\newcommand{\eps}{\varepsilon}

\newcommand{\be}{\begin{equation}}
\newcommand{\ee}{\end{equation}}
\newcommand{\bea}{\begin{eqnarray}}
\newcommand{\eea}{\end{eqnarray}}
\newcommand{\ba}{\begin{array}}
\newcommand{\ea}{\end{array}}

\newcommand{\und}{\qquad\textrm{and}\qquad}
\newcommand{\nn}{\nonumber}

\renewcommand{\=}{\ =\ }
\newcommand{\diff}{\textrm{d}}
\newcommand{\sfrac}[2]{{\textstyle\frac{#1}{#2}}}

\begin{document}
\thispagestyle{empty}
\vspace{2cm}
\begin{flushright}
\end{flushright}\vspace{2cm}
\begin{center}
{\LARGE \bf  ${\cal N}{=}4$ Supersymmetry and the BPST Instanton}
\end{center}
\vspace{1cm}

\begin{center}
{\Large\bf  S.~Krivonos${}^{a}$, O.~Lechtenfeld${}^{b}$ and A.~Sutulin${}^{a}$ }
\end{center}

\begin{center}
${}^a$ {\it
Bogoliubov  Laboratory of Theoretical Physics, JINR,
141980 Dubna, Russia} \vspace{0.2cm}

${}^b$ {\it Leibniz Universit\"at Hannover,
Appelstr. 2, D-30167 Hannover, Germany}\vspace{0.2cm}

\end{center}
\vspace{2cm}

\begin{abstract}
\noindent In this paper we construct the Lagrangian and Hamiltonian
formulations of ${\cal N}{=}4$ supersymmetric systems describing the motion
of an isospin particle on a conformally flat four-manifold with SO(4) isometry
carrying the non-Abelian field of a BPST instanton.
The conformal factor can be specified to yield various particular systems,
such as superconformally invariant mechanics as well as a particle on the
four-sphere, the pseudosphere or on $\mathbb{R}\times\mathbb{S}^3$.
The isospin degrees of freedom arise as bosonic components of
an additional fermionic ${\cal N}{=}4$ supermultiplet,
whose other components are rendered auxiliary by a nonlocal redefinition.
Our on-shell component action coincides with the one recently proposed
in {\tt arXiv:0912.3289}.
\end{abstract}
\newpage

\setcounter{page}{1}
\setcounter{equation}0
\section{Introduction}
In the past decade a lot of attention was paid to the higher-dimensional
quantum Hall effect. In many respects, the four-dimensional Hall effect,
as formulated by Zhang and Hu~\cite{hall}, was a breakthrough result.
At the heart of their approach was the second Hopf map, and most subsequent
developments utilized and extended this idea. Among the results, we mention
an eight-dimensional variant of the Hall effect~\cite{ber},
an extension of the quantum Hall systems to $\mathbb{CP}$~manifolds~\cite{CP}
and hyperbolic versions of the quantum Hall effect~\cite{Hasebe}.

Another activity concerns  the extensions of the quantum Hall effect to
supersymmetric theories. From a formal point of view, such an extension
requires a supersymmetric mechanics of an isospin-carrying particle moving
in the background of magnetic monopoles.
By now it is well known \cite{{FIO1},{Braz},{BK},{FIO},{KO},{BKS1}}
that to invent monopole-type interactions in Lagrangian mechanics one has
to involve ``isospin" variables with a specific kinetic energy of first order
in the time derivatives. In supersymmetric systems these ``isospin'' variables
become part of some supermultiplet, whose spinor components are auxiliary.
The first realization of this idea was proposed in~\cite{FIO1}, where isospin
bosonic and auxiliary fermionic degrees of freedom constitute an auxiliary
gauge supermultiplet. More recently, \cite{Smilga}~has demonstrated that
the minimal coupling to an external non-Abelian self-dual background
works perfectly in the case of four-dimensional ${\cal N}{=}4$ supersymmetric
mechanics, and in~\cite{IKS1} the Lagrangian formulation has been given by
the use of harmonic superspace.

Here, we present an alternative approach, which utilizes ordinary superspace
together with a nonlocal component redefinition. This procedure was developed
in~\cite{BK} and has been applied to three-dimensional supersymmetric mechanics
in~\cite{BKS1}. The main idea of this approach is the replacement of physical
fermions by auxiliary ones, which we describe as follows.
Suppose we have at hands a $(4,4,0)$ fermionic supermultiplet $\Psi^\alpha$
with four physical fermions $\{\psi^\alpha,\bpsi_\alpha\}$ and four auxiliary
bosons $\{ v^i,{\bar v}_i\}$, subject to the standard $d{=}1$, ${\cal N}{=}4$
Poincar\'{e} supersymmetry transformations
\be\label{ad1}
\delta \psi^1=-\bar\epsilon{}^i {\bar v}_i,\; \delta\psi^2=\epsilon_i{\bar v}{}^i,\quad \delta v^i=-2i\epsilon^i\dot\bpsi{}^1+2i\bar\epsilon{}^i\dot\bpsi{}^2,\;
\delta {\bar v}_i=-2i\epsilon_i\dot\psi{}^1+2i\bar\epsilon_i\dot\psi{}^2.
\ee
If we make the formal replacement
\be\label{dualiz}
\dot{\psi}{}^\alpha\ \to\ \chi^\alpha \qquad\textrm{and}\qquad
\dot{\bpsi}_\alpha\ \to\ \bchi_\alpha,
\ee
we get a new supermultiplet ${\cal V}{}^i$ of $(0,4,4)$ type with
components $\{ v^i, {\bar v}_i, \chi^\alpha, \bchi_\alpha\}$, which
transform as
\be\label{ad2}
\delta \chi^1=-\bar\epsilon{}^i \dot{\bar v}_i,\;
\delta\chi^2=\epsilon_i\dot{\bar v}{}^i,\quad \delta
v^i=-2i\epsilon^i\bchi{}^1+2i\bar\epsilon{}^i\bchi{}^2,\; \delta
{\bar v}_i=-2i\epsilon_i\bchi{}^1+2i\bar\epsilon_i\chi{}^2.
\ee
The goal is to construct an ${\cal N}{=}4$ supersymmetric Lagrangian for
the components of~${\cal V}{}^i$. To this end, we couple $\Psi^\alpha$
to some ``matter'' multiplet~$Q$ in a standard superspace action
$S[\Psi^\alpha,Q]$.
If we make sure that the fermionic components~$\psi^\alpha$ are cyclic in
this action, i.e.~they appear only via their derivatives~$\dot{\psi}^\alpha$,
then we may perform the replacement~\p{dualiz} on the component level and
obtain a supersymmetric and local action for the fields appearing in
${\cal V}{}^i$ and~$Q$. To be sure, such a replacement alters the dynamics:
In terms of $\Psi^\alpha$, it amounts to putting to zero the momentum
canonically conjugate to~$\psi^\alpha$. However, we are not interested in
the dynamics of~$S[\Psi^\alpha,Q]$ but in the physics of the new action
governing the dynamics of the components of ${\cal V}{}^i$ and~$Q$.

In the present paper, we choose for the matter~$Q$ a one-dimensional
hypermultiplet $Q^{ia}$~\cite{hyper1,hyper2,hyper3,hyper4,hyper5}
and couple it minimally to~$\Psi^\alpha$,
\be\label{Sinit0}
S[\Psi^\alpha,Q^{ia}]\= \int\!\diff^4\theta\,\diff t\;\left[
F(Q)\ +Y(Q) \Psi^\alpha \bar\Psi_\alpha \right] .
\ee
The condition of cyclicity of $\psi^\alpha$ in this action restricts the
function $Y(Q)$ to be harmonic,
\be\label{H}
\frac{\partial^2}{\partial Q^{ia} \partial Q_{ia}} Y=0.
\ee
This generalizes to four dimensions the cases of one- and three-dimensional
${\cal N}{=}4$ supersymmetric mechanics with isospin variables considered
in~\cite{{BK},{BKS1}}.
The simple action \p{Sinit0} will lead to a minimal coupling to the instanton
if we choose the $SO(4)$ invariant solution of the condition \p{H} as
\be\label{SOL}
Y=\rho + \frac{2}{Q^{ia}Q_{ia}} \qquad\textrm{for}\quad Q^{ia}\neq0.
\ee
Later on, the constant $|\rho|$ becomes the size of the instanton.\footnote{
In four dimensions this constant plays an essential role, in contrast to the
three-dimensional case \cite{BKS1}.}
To have an $SO(4)$ invariant system, we also restrict the arbitrary
function~$F(Q)$ to depend only on the SO(4)~invariant combination
\be\label{F}
X=2/(Q^{ia}Q_{ia})
\ee
of the hypermultiplet fields $Q^{ia}$.
Thus, we arrive at the same action as proposed in~\cite{BK},
\be\label{Sinit}
S[\Psi^\alpha,X] \= \int\!\diff^4\theta\,\diff t\;\left[
F(X)\ +\ \left( X{+}\rho\right) \Psi^\alpha \bar\Psi_\alpha \right].
\ee
The matter has component content $X=\{x, A^{(ij)},\eta^i, \bar\eta_i\}$,
which transforms as
\be\label{Tr1}
\delta x=-i\epsilon_i\eta^i-i\bar\epsilon{}^i\bar\eta_i,\quad
\delta\eta{}^i=-\bar\epsilon{}^i{\dot x}-i\bar\epsilon{}^j
A^i_j,\; \delta \bar\eta_i=-\epsilon_i{\dot x}+i\epsilon_j
A_i^j,\quad \delta A_{ij} = -\epsilon_{(i}\dot\eta_{j)} +
\bar\epsilon_{(i}\dot{\bar\eta}{}_{j)}.
\ee
We stress that it is the composite structure~\p{F} of the superfield $X$
which causes our particle to interact with the instanton. If instead
we treat $X$ as an independent ${\cal N}{=}4$~superfield, the isospin degrees
of freedom will decouple and the resulting system will describe a
particle in the field of a Dirac monopole~\cite{FIO,KO}. On the other hand,
employing the composite-field concept in three dimensions produces a coupling
to the Wu-Yang monopole~\cite{BKS1}.

At this point we perform the integration over the $\theta$'s and then apply
our replacement recipe~\p{dualiz}.
A straightforward computation yields the ${\cal N}{=}4$ supersymmetric
off-shell component action~\cite{BK}
\bea\label{Sa}
S&=&\int\!\diff t\left[\frac{1}{8}G {\dot x}{}^2-\frac{1}{16} G
A^{ij}A_{ij}+\frac{i}{8} G\left(\dot\eta{}^i\bar\eta_i-
\eta^i\dot{\bar\eta}_i\right)+\frac{1}{8}G'\eta^i\bar\eta{}^jA_{ij}-\frac{1}{32}G''\eta^i\eta_i\bar\eta_j\bar\eta{}^j\right. \nn\\
&& -(x+\rho) \left({ \chi}{}^1{\bchi}{}^2-{ \chi}{}^2{\bchi}{}^1\right)+\frac{i}{4} (x+\rho) \left( {\dot v}_i {\bar v}{}^i-
v_i\dot{\bar v}{}^i\right)+\frac{1}{4}A_{ij}v^i{\bar v}{}^j\nn\\
&& \left. +\frac{1}{2}\eta_i\left({\bar v}{}^i \bchi{}^2+v^i
\chi{}^2\right)+\frac{1}{2}\bar\eta{}^i\left(v_i \chi{}^1+{\bar
v}_i\bchi{}^1\right) \right],
\eea
which describes the interaction of eight bosons
$\left\{ x, A^{(ij)}, v^i, {\bar v}_i\right\}$ and eight fermions
$\left\{ \eta^i, \bar\eta_i, \chi^\alpha,\bchi_\alpha\right\}$
living on the one-dimensional worldline of a particle.
Here, $G=F''(x)$ is an arbitrary function depending on $x$ only,
$\rho$ is a free parameter, and all indices run over 1 and~2.
This action is our starting point.

In the following section we perform several changes of variables and eliminate
auxiliary ones, in order to bring out explicitly the instanton coupling.
In Section~3 we present the supercharges and the Hamiltonian, as well as the
four-dimensional translation and rotation generators. The configuration-space
metric of our system is SO(4)-invariant and conformally flat,
thus depends only on the single `radial' function $G(x)$.
In Section~4 we specialize this metric to obtain a few interesting examples,
such as a particle on the sphere $\mathbb{S}{}^4$ interacting with a
BPST~instanton located in its center. Finally, in the Conclusions
we shortly discuss the bosonic SO(5)~symmetry which naturally appears
in the latter case and which is explicitly broken by fermionic terms.

\setcounter{equation}0
\section{The instanton coupling}

In our model the four-dimensional nature of the theory is encoded in the composite structure of
the superfield $X$ \p{F}. The net effect of such a representation is summarized in the composite
structure of the "auxiliary" components $A^{ij}$, which are now
expressed via the components of $Q^{ia}$ as
\be\label{A}
A_{ij}=\-i x \left( {\dot q}{}^a_i q_{ja} + {\dot q}{}^a_j q_{ia}\right)
-\frac{1}{x} \left( \eta_i \bar\eta_j+\eta_j \bar\eta_i\right)\ .
\ee
Here, we have used a polar representation of the bosonic $Q^{ia}$-components,
\be\label{q}
Q^{ia}Q_{ia}| \= \frac{2}{X|} \ =:\ \frac2x \und
Q^{ia}| \ =:\ \frac{q^{ia}}{\sqrt{x}}
\qquad \Rightarrow \qquad q^{ia}q_{ia}=2\ .
\ee

We substitute the expression~\p{A} for $A_{ij}$ into the component action
\p{Sa} and eliminate the auxiliary fermions $\chi^\alpha$ and $\bchi_\alpha$
by their equations of motion, obtaining
\bea\label{Sb}
S&=&\int\!\diff t\;\left\{
\frac{1}{8} G\left[ {\dot x}{}^2 +\frac{x^2}{2} \omega^{ij}\omega_{ij}
+i\left( \dot\eta{}^i \bar\eta_i -\eta^i\dot{\bar\eta}_i\right)\right]
-\frac{i}{8}\left( 2G +x G'\right) \omega_{ij} \eta^i \bar\eta{}^j
-\frac{x^2 G''+6 x G'+6G}{32x^2} \eta^2 \bar\eta{}^2 \right.\nn \\
&&\left.  +\frac{i}{4}\; (x+\rho)\;
\left( {\dot v}_i {\bar v}{}^i -v_i \dot{\bar v}{}^i \right)
-\frac{i}{4}\; x\; \omega_{ij} v^i {\bar v}{}^j
-\frac{\rho}{4x(x{+}\rho)} v^i {\bar v}{}^j
\left(\eta_i \bar\eta{}_j+\eta_j \bar\eta{}_i\right)\right\}\ ,
\eea
where
\be\label{w}
\omega_{ij}\={\dot q}{}^a_i q_{ja}+{\dot q}{}^a_j q_{ia}\ .
\ee
The action \p{Sb} describes four physical bosons
$\{ x, q^{ia}:q^{ia}q_{ia}{=}2\}$, four physical fermions
$\{ \eta^i, \bar\eta_i\}$, and four ``isospin'' variables
$\{v^i, {\bar v}_i \}$.

The variables we used until now were rather useful for discussing
${\cal N}{=}4$ supersymmetry properties. However, for clarifying the
interactions disguised in~\p{Sb} it is preferable to change variables.
We do this in two steps.

First, in order to simplify the kinetic terms for all variables,
we rescale them to
\be\label{nb1}
Y^{ia} \= \sqrt{\frac{x}{2}}\,q^{ia}
\quad\Rightarrow\quad Y^{ia}Y_{ia}=x\ , \qquad
u^i\=\sqrt{Y^2{+}\rho}\;v^i\ ,\qquad \xi^i\=\frac{\sqrt{G}}{2}\,\eta^i\ .
\ee
In addition, we introduce the isospin and fermionic spin currents
as useful bilinears:
\be\label{IS}
I^A \= \frac{i}{2} \left( \sigma^A\right)_i^j\, u^i\, {\bar u}_j\ , \qquad
\Sigma^A\= -i \left( \sigma^A\right)_i^j\, \xi^i\,\bar\xi_j\ , \qquad
\textrm{with}\quad A=1,2,3\ ,
\ee
where the $\sigma^A$-matrices are normalized as \
$[\sigma^A,\sigma^B]=2i\epsilon^{ABC}\sigma^C$.
In terms of these variables  the action \p{Sb} reads
\bea\label{Sc}
S&=&\int\!\diff t\;\left\{\frac{1}{2}\,G\,Y^2 \,{\dot Y}{}^{ia}{\dot Y}_{ia}
+\frac{i}{2} \left( \dot\xi{}^i \bar\xi_i -\xi^i\dot{\bar\xi}_i\right)-
\frac{i}{4}\; \left( {\dot u}^i {\bar u}{}_i -u^i \dot{\bar u}{}_i \right)-
\frac{1}{2\,Y^2\, G}\left( 2G +Y^2 G'\right) \Omega^A\,\Sigma^A \right. \nn\\
&&\left.  +\frac{1}{2\,(Y^2{+}\rho)}\; \Omega^A\,I^A-
\frac{Y^4 G''+6 Y^2 G'+6G}{3\,Y^4\, G^2} \Sigma^A \Sigma^A
+\frac{2\rho}{G\,Y^2\,(Y^2{+}\rho)^2} I^A \Sigma^A\right\}\ ,
\eea
where we introduced
\be\label{Omega}
\Omega^A \= \bigl({\dot Y}{}^{ja} Y_{ia}+Y^{ja}{\dot Y}{}_{ia}\bigr)
\left(\sigma^A\right)_j^i\ .
\ee

Second, we pass to four-dimensional vector coordinates $y^\mu$ via
\be\label{Y}
Y^{ia}\=\frac{1}{\sqrt{2}} \eps^{ik} y^{\mu} (\sigma^{\mu})^a_k  \qquad
\Rightarrow \qquad Y^{ia} Y_{ia} =  y^{\mu} y^{\mu}\ ,
\ee
where the four sigma-matrices are defined as \
$\sigma^{\mu}=(i\sigma^A,\mathbf{1})$ \ with \ $A=1,2,3$ \ and \ $\mu=1,2,3,4$.
It is easy to check that the ingredient $\Omega^A$ in~\p{Sc}
acquires the nice form
\be
\Omega^A \= 2i\,\eta^A_{\mu\nu}\, y^\mu {\dot y}{}^\nu
\ee
involving the self-dual t'Hooft symbol
\be
\eta^A_{\mu\nu}\=\delta^A_\mu \delta_{\nu 4}-\delta^A_\nu \delta_{\mu 4}
+ \epsilon^A_{\; \;\mu\nu 4} \qquad\Rightarrow\qquad
(\delta_{\mu\rho}\delta_{\nu\sigma}-\sfrac12\epsilon_{\mu\nu\rho\sigma})\,
\eta^A_{\rho\sigma}\=0\ .
\ee
Combining everything we get the final form of the action,
\bea\label{finaction}
S &=& \int\!\diff t\; \left[ \frac{g}{2}\; \dot y^{\mu} \dot y^{\mu} +
\frac{i}{2}\, \left( \dot\xi{}^i \bar\xi_i -\xi^i\dot{\bar\xi}_i\right)
-\frac{i}{4}\, \left( {\dot u}{}^i {\bar u}_i -u^i \dot{\bar u}_i \right)
\right. 
- \frac{i}{y^2\,g}\left( g+y^2\,g'\right)\eta^A_{\mu\nu} y^{\mu}\dot y^{\nu}\,
\Sigma^A
\nn \\
&& +\frac{i}{y^2{+}\rho}\,\eta^A_{\mu \nu} y^{\mu} \dot y^{\nu}\,I^A
 \left.
+ \frac{2\rho}{(y^2{+}\rho)^2\, g}\,I^A \Sigma^A
-\frac{1}{3\, y^2\,g^2} \left( 2g+4 y^2 g'+ y^4\, g''\right)\Sigma^A \Sigma^A
\right] ,
\eea
where the metric function $g$ is defined as
\be
g(y^2) \= y^2\, G(y^2)\ .
\ee

The action \p{finaction} is our main result. It describes ${\cal N}{=}4$
supersymmetric four-dimensional isospin particles moving in the field of
a BPST instanton. Indeed, from~\p{finaction} one sees that the bosonic part
of the vector potential reads
\be\label{YangP}
{\cal A}{}_\mu \= -\frac{i}{y^2{+}\rho}\,\eta^A_{\mu \nu}\,y^{\nu}\,I^A
\qquad\Rightarrow\qquad
{\cal F}_{\mu \nu}\ \equiv\ \partial_\mu {\cal A}_\nu-
\partial_\nu {\cal A}_\mu+ \left[ {\cal A}_\mu , {\cal A}_\nu\right] \=
\frac{2 i\, \rho\, \eta^A_{\mu \nu}  \,I^A}{(y^2{+}\rho)^2}\ ,
\ee
which is of the familiar instanton form if we may view $I^A$, as defined
in~\p{IS}, as proper isospin matrices.\footnote{
For a solution to the Yang-Mills equations we must have
$\left[ I^A, I^B\right]=2i \epsilon^{ABC}I^C$.}

The on-shell component action \p{finaction} coincides
(modulo some redefinitions) with the one constructed recently within
the harmonic superspace approach in~\cite{IKS1}.
Surely, the most general case of the action \p{Sinit0}
with an arbitrary prepotential $F(Q)$
and a more general harmonic function $Y(Q)$ could be easily considered.

To close this Section, let us comment on the appearance of the t'Hooft symbol
in our construction. The definition of $\Omega^A$ in~\p{Omega} makes use
of the $su(2)$ algebra generated by the $\sigma^A$, which gets embedded into
the self-dual part of the $so(4)$ symmetry group via~\p{Y}.
If instead we embed into the anti-self-dual part, by replacing
$\sigma^\mu$ with ${\bar\sigma}{}^\mu=(-i\sigma^A,\mathbf{1})$,
we shall arrive at
\be
\Omega^A \= 2i\,\bar\eta{}^A_{\mu\nu}\, y^\mu {\dot y}{}^\nu
\qquad\textrm{with}\qquad
\bar\eta{}^A_{\mu\nu}=-\delta^A_\mu \delta_{\nu 4}+\delta^A_\nu \delta_{\mu 4}
+ \epsilon^A_{\; \;\mu\nu 4}\ ,
\ee
and the vector potential becomes
\be\label{YangPD}
{\cal A}{}_\mu \= -\frac{i}{y^2{+}\rho}\,\bar\eta{}^A_{\mu \nu} y^{\nu}\,I^A
\qquad\Rightarrow\qquad
(\delta_{\mu\rho}\delta_{\nu\sigma}+\sfrac12\epsilon_{\mu\nu\rho\sigma})\,
{\cal F}_{\rho\sigma}\=0\ ,
\ee
producing the BPST anti-instanton.

\setcounter{equation}0
\section{Hamiltonian and Supercharges}
In order to find the classical Hamiltonian, we follow the standard procedure
for quantizing a system with bosonic and fermionic degrees of freedom.
{}From the action~\p{finaction} we define the momenta
$(P_\mu, \pi_i, \bar\pi{}^i, p_i,{\bar p}{}^i)$ conjugated to
$(y^\mu, \xi^i, \bxi_i, u^i, {\bar u}_i)$ as
\bea\label{momenta}
&&
P_\mu \=
g\,{\dot y}{}_\mu\ -\ i\eta^A_{\mu\nu} y^\nu\left(\frac{1}{y^2{+}\rho}I^A
-\frac{g{+}y^2 g'}{y^2 g} \Sigma^A\right)\,, \nn \\
&&
\pi_i = \frac{i}{2}\, \bar \xi_i\ , \quad
\bar \pi^i = \frac{i}{2}\, \xi^i\ , \quad
p_i = -\frac{i}{4}\, {\bar u}_i\ , \quad
\bar p^i = \frac{i}{4}\, u^i\ ,
\eea
and introduce Dirac brackets for the canonical variables,
\be\label{PB}
\{y^\mu,P_\nu \} \= \delta^{\mu}_{\nu}\ , \qquad
\{ \xi^i, \bar \xi_j \} \= i \delta^i_j\ , \qquad
\{ u^i, {\bar u}_j \} \= 2i \delta^i_j\ .
\ee
As usual, the canonical momenta $P^\mu$ differ by the vector-potential shift
from the kinematical momenta
\be\label{phat}
\widehat{P}_\mu\ :=\ g\,{\dot y}{}_\mu\=
P_\mu\ +\ i \eta^A_{\mu\nu} y^\nu\left(\frac{1}{y^2+\rho} I^A
-\frac{g+y^2 g'}{y^2 g} \Sigma^A\right)\ ,
\ee
whose Dirac brackets contain the instanton field strength,
\bea\label{PP}
\left\{ \widehat{P}_\mu,\widehat{P}_\nu \right\}
&=& \frac{2i\rho}{(y^2{+}\rho)^2} \eta^A_{\mu\nu} I^A\ -\
2i \frac{2 g^2+3 y^2 g g'+y^4 (g')^2 }{y^2 g^2} \eta^A_{\mu\nu} \Sigma^A \nn\\
&& -\ 2i \frac{2 g^2 +2 y^4 (g')^2  +2 y^2 g g'- y^4 g g''}{y^4 g^2}
\left( y_{\mu}\,\eta^A_{\nu\rho}-y_{\nu}\,\eta^A_{\mu\rho}\right)y^\rho \,
\Sigma^A\ .
\eea
One may check that the supercharges
\bea\label{QQ}
Q^i \!\!&=&\!\! \frac{i}{ \sqrt{y^2\, g}} \Big (\delta^i_j\,\delta_{\mu\nu}
\ -\ i\eta^A_{\mu\nu}\,(\sigma^A)^i_j \,\Big)\,y_{\mu}\, P_{\nu}\,\bar \xi^j
\ -\ \frac{i\sqrt{y^2}}{(y^2{+}\rho)\sqrt g}\,(\sigma^A)^i_j \,\bar \xi^j I^A
\ -\ i\frac{g-y^2 g'}{3 \sqrt{y^2}\,  g \sqrt g}  (\sigma^A)^i_j \, \bar \xi^j
\Sigma^A \nn\\
&=&\!\! \frac{i}{\sqrt{y^2\,g}}\Big(\delta^i_j\,\delta_{\mu\nu}
\ -\ i\eta^A_{\mu\nu}\,(\sigma^A)^i_j\,\Big)\,y_{\mu}\,\widehat{P}{}_\nu\,\
\bar\xi^j
\ -\ \frac{2i}{3}\left(\frac{2 g +y^2 g'}{\sqrt{y^2}\,g\sqrt g}\right)
(\sigma^A)^i_j \, \bar \xi^j \Sigma^A \ ,\nn\\
\bar Q_i \!\!&=&\!\! \frac{i}{\sqrt{y^2\, g}} \Big (\delta^j_i\,\delta_{\mu\nu}
\ + \ i \eta^A_{\mu \nu}\, (\sigma^A)^j_i \Big )\, y_{\mu}\, P_{\nu} \, \xi_j
\ + \ \frac{i \sqrt{y^2}}{(y^2{+}\rho) \sqrt g}\, (\sigma^A)^j_i \, \xi_j I^A
\ +\ i\frac{g-y^2 g'}{3 \sqrt{y^2}\,g\sqrt g}(\sigma^A)^j_i\,\xi_j\Sigma^A\nn\\
&=&\!\! \frac{i}{\sqrt{y^2\, g}} \Big (\delta^j_i\,\delta_{\mu\nu}
\ +\ i\eta^A_{\mu\nu}\,(\sigma^A)^j_i \Big)\,y_{\mu}\,\widehat{P}{}_\nu \,\xi_j
\ +\ \frac{2i}{3}\left(\frac{2 g+ y^2 g'}{\sqrt{y^2}\,g\sqrt g}\right)
(\sigma^A)^j_i \, \xi_j \Sigma^A
\eea
and the Hamiltonian
\be\label{HAM}
H \= \frac{1}{2\, g}\,{\widehat P}^{\mu}{\widehat P}^{\mu}
\ -\ \frac{2\rho}{(y^2{+}\rho)^2 g}\,I^A \Sigma^A
\ +\ \frac{1}{3\,y^2\,g^2}\Big(2g+ 4 y^2\, g' + y^4\, g''\Big)\Sigma^A\Sigma^A
\ee
form the standard ${\cal N}{=}4, d=1$  Poincar\`{e} superalgebra
\be\label{poincare}
\left\{ Q^i,{\bar Q}_j\right\} \= 2 i\,\delta^i_j\,H \ ,\qquad
\left[ Q^i, H \right] \= \left[ {\bar Q}_j, H \right] \= 0\ .
\ee

The spin variables $u^i,\bar{u}_i$ enter the Hamiltonian only through the
isospin currents $I^A$, which commute with everything, excluding themselves:
\be\label{S}
\left\{ I^A, I^B\right\} \= 2i \epsilon^{ABC} I^C
\und \left\{ I^A I^A, H \right\} \= 0,
\ee
just forming  an $su(2)$ algebra with respect to the brackets \p{PB}. Thus,
$I^A$ may be interpreted as classical isospin matrices at fixed isospin~$I$.
Analogously, also the fermions appear in the Hamiltonian only through
the combination $\Sigma^A$ which likewise obeys
\be\label{Sigma}
\left\{ \Sigma^A, \Sigma^B\right\} \= 2i \epsilon^{ABC} \Sigma^C
\und \left\{ \Sigma^A \Sigma^A, H \right\} \= 0\ ,
\ee
thus providing a description for the fermionic spin degrees of freedom.
Clearly, the SO(4) invariance of our system is realized in a standard way
through the generators
\be\label{so4}
M_{\mu\nu}\= y_\mu P_\nu-y_\nu P_\mu
\ -\ \frac{i}{2} \eta^A_{\mu\nu}\left( I^A +\Sigma^A \right)
\qquad\Rightarrow\qquad \left\{M_{\mu\nu},H \right\}\=0\ .
\ee
Thus, we conclude that the Hamiltonian \p{HAM} indeed describes the motion of
an ${\cal N}{=}4$ supersymmetric isospin particle in an BPST instanton background.

The supercharges \p{QQ} and Hamiltonian \p{HAM} have the same structure as
those ones presented in~\cite{Smilga}. Therefore, our component action~\p{Sa}
provides a formulation alternative to the one of~\cite{IKS1}.

\setcounter{equation}0
\section{Cases of special interest}
So far our consideration was general, and the function $g(y^2)$
in~\p{finaction} was arbitrary. Let us now specify it to produce some cases
of particular interest.

\subsection{Superconformal invariant models}
Invariance under the most general ${\cal N}{=}4$ superconformal group
$D(2,1;\alpha)$ in one dimension~\cite{sorba} is achieved for the
choice~\cite{IKL1}
\footnote{${\cal N}{=}4$ $D(2,1;\alpha)$ superconformal mechanics 
without ``isospin" degrees of freedom has been constructed in \cite{Mich1}.}
\be\label{SConf}
g(y^2)\=(y^2)^{-1-\frac{1}{\alpha}} \qquad\textrm{for}\quad \alpha\neq0\ .
\ee
In addition, superconformal invariance demands fixing our parameter
to $\rho=0$, so the instanton must have zero size.
Of special interest are the subcases $\alpha=-1$ and $\alpha=1$
corresponding to the SU$(1,1|2)$ and OSp$(4|2)$ superconformal groups,
for which the metric is flat (with $z^\mu=y^{-2}y^\mu$ in the second case):
\bea\label{minus1}
S_{\alpha=-1}\!\!&=&\!\! \int\!\diff t\; \left[
\frac{1}{2} \, \dot y^{\mu} \dot y^{\mu} + \frac{i}{2}\,
\left( \dot\xi{}^i \bar\xi_i{-}\xi^i\dot{\bar\xi}_i\right)
-\frac{i}{4}\, \left( {\dot u}{}^i {\bar u}_i{-}u^i \dot{\bar u}_i \right)
+\frac{i}{y^2}\,\eta^A_{\mu\nu}y^{\mu}\dot y^{\nu}\,\left(I^A{-}\Sigma^A\right)
-\frac{2}{3\, y^2} \Sigma^A \Sigma^A \right]\ , \nn \\
S_{\alpha=1}\!\!&=&\!\! \int\!\diff t\; \left[
\frac{1}{2}\; \dot z^{\mu} \dot z^{\mu} + \frac{i}{2}\,
\left( \dot\xi{}^i \bar\xi_i{-}\xi^i\dot{\bar\xi}_i\right)
-\frac{i}{4}\, \left( {\dot u}{}^i {\bar u}_i{-}u^i \dot{\bar u}_i \right)
+\frac{i}{z^2}\,\eta^A_{\mu\nu}z^{\mu}\dot z^{\nu}\,\left(I^A{+}\Sigma^A\right)
\right]\ .
\eea
As expected, the four-fermion term disappeared in the
$OSp(4|2)$ invariant case.

\subsection{$\mathbb{R}\times \mathbb{S}^3$ case}
In the limit of $\alpha\to\infty$ we obtain another special case,
\be
g(y^2)\=(y^2)^{-1}\ .
\ee
With this choice, the kinetic term for the $y^\mu$ variables in
the action~\p{finaction} acquires the form
\be
\frac{1}{2 y^2}\, \dot y^\mu\, \dot y_\mu \=
\frac{1}{2}\left(\frac{{\dot{\tilde y}}^2}{{\tilde y}{}^2} +
{\dot{\hat y}}{}^\mu\,{\dot{\hat y}}{}^\mu \right)
\qquad\textrm{with}\qquad
y^\mu={\tilde y}{\hat y}{}^\mu, \quad {\hat y}{}^\mu {\hat y}{}^\mu=1\ ,
\ee
and thus we meet an $\mathbb{R}\times \mathbb{S}^3$ geometry in the
bosonic sector. The full action then reads
\be\label{RS3}
S_{\alpha\to\infty} \= \int\!\diff t\;\left[
\frac{1}{2 y^2} \dot y^{\mu} \dot y^{\mu}
+ \frac{i}{2}\, \left( \dot\xi{}^i \bar\xi_i {-}\xi^i\dot{\bar\xi}_i\right)
-\frac{i}{4}\, \left( {\dot u}{}^i {\bar u}_i {-}u^i \dot{\bar u}_i \right)
+\frac{i}{y^2{+}\rho}\,\eta^A_{\mu \nu} y^{\mu} \dot y^{\nu}\,I^A
+ \frac{2\rho y^2}{(y^2{+}\rho)^2}\,I^A \Sigma^A \right] .
\ee

\subsection{Sphere $\mathbb{S}^4$ and pseudo-sphere cases}
To describe the sphere $\mathbb{S}^4$ or the pseudo-sphere one has to choose
\be\label{s4p}
g \= \frac{1}{(\rho+y^2)^2}\ ,
\ee
with $\rho>0$ for the sphere or $\rho<0$ for the pseudo-sphere.
The corresponding action becomes
\bea\label{s4}
S_{\mathbb{S}^4} \!\!&=&\!\! \int\!\diff t\;\left[
\frac{1}{2\,(\rho{+} y^2)^2}\; \dot y^{\mu} \dot y^{\mu}
+ \frac{i}{2}\, \left( \dot\xi{}^i \bar\xi_i{-}\xi^i\dot{\bar\xi}_i\right)
-\frac{i}{4}\, \left( {\dot u}{}^i {\bar u}_i{-}u^i \dot{\bar u}_i \right)
\right. \nn \\
&&\left. +\ \frac{i}{y^2{+}\rho}\,\eta^A_{\mu \nu} y^{\mu} \dot y^{\nu}\,I^A
-i\frac{\rho-y^2}{y^2(\rho{+}y^2)}\eta^A_{\mu\nu}y^{\mu}\dot y^{\nu}\,\Sigma^A
+ 2\rho\,I^A\Sigma^A+\frac{2\rho(2y^2{-}\rho)}{3y^2}\Sigma^A\Sigma^A\right]\ .
\eea

\subsection{Very simple system}
Rather than specializing to a simple bosonic manifold like we did so far,
one might try to simplify the fermionic sector instead.
Here, the maximal simplification occurs for
\be\label{strange1}
g\=\frac{1}{y^2\,(y^2+\rho)}\ .
\ee
With this choice, the system possesses an additional conserved current,
\be\label{W}
W\= I^A \Sigma ^A\ .
\ee
In addition, the kinematical momenta simplify to
\be\label{strange2}
\widehat{P}{}_\mu\= P_\mu\ +\ i\eta^A_{\mu\nu}y^\nu\frac{1}{y^2{+}\rho}
\left( I^A + \Sigma^A\right)
\ee
and, therefore,
\be\label{strange3}
\left\{ \widehat{P}_\mu,\widehat{P}_\nu \right\} \=
\frac{2i\,\rho}{(y^2{+}\rho)^2}\,\eta^A_{\mu\nu}\,(I^A+\Sigma^A) \=
{\cal F}_{\mu\nu}\big|_{I\to I+\Sigma}\ .
\ee
Finally, the `very simple' Hamiltonian reads
\be\label{strange4}
H \= \frac{y^2 (y^2{+}\rho)}{2}\, {\widehat P}_{\mu} {\widehat P}_{\mu}
\ -\ \frac{2\rho y^2}{y^2{+}\rho}\,\left( I^A \Sigma^A +\frac{1}{3}
\Sigma^A \Sigma^A \right)\ .
\ee

\setcounter{equation}0
\section{Conclusion}
In this paper we have constructed the Lagrangian and Hamiltonian formulations
of ${\cal N}{=}4$ supersymmetric systems describing the motion of isospin
particles on a conformally flat four-manifold with a BPST SU(2)~instanton.
Due to SO(4) rotation invariance around the instanton location,
the conformal factor depends on the radial variable only.
It was further specified to capture some particularly interesting systems,
including superconformally invariant mechanics, a particle living on
the sphere, the pseudo-sphere, or on $\mathbb{R}\times\mathbb{S}^3$.
The isospin variables entered the action as the bosonic components of an
auxiliary ${\cal N}{=}4$ supermultiplet. Its other components were auxiliary
and became expressed on-shell through the physical fermions.
It is obvious how to generalize the action~\p{Sinit0} to an arbitrary
prepotential~$F(Q)$ and a general harmonic function~$Y(Q)$.

Starting from the off-shell component action~\p{Sa}~\cite{BK}
we derived the action~\p{finaction}, which coincides with the one proposed
recently in~\cite{IKS1}. It is a matter of taste to prefer one approach over
another. What makes us enthusiastic about the present construction
is the extreme simplicity of the precursor action~\p{Sinit0} in ordinary
superspace. Of course, one must revert to the component level for applying
the nonlocal replacement recipe~\p{dualiz}.

The sphere case, treated in Subsection~4.3, is of enhanced interest due to
its possible relation with the four-dimensional Hall effect.
Indeed, our system provides a new ${\cal N}{=}4$ supersymmetric extension of
the model considered by Zhang and Hu~\cite{hall}, with which it coincides
in the bosonic sector. It is tempting to investigate the spectrum of our
system and to analyze the role played by the supersymmetry of~\p{s4}.

Unfortunately, not everything looks nice in our system.
First of all, the implicit SO(5) symmetry of the four-dimensional Hall effect,
which played a crucial role in the computation of spectra in~\cite{hall},
is explicitly broken by ${\cal N}{=}4$ supersymmetry.
Indeed, to be the same constant at all points of the four-sphere,
the right-hand side of~\p{PP} can only depend on~$y^2$.
This necessary condition results in an equation
for the configuration-space metric~$g$,
\be\label{so5}
2 g^2 +2 y^4 (g')^2  +2 y^2 g g'- y^4 g g''\=0
\qquad \Rightarrow \qquad g\=\frac{c_1}{y^2 (c_2 y^2{+}\rho)}\ .
\ee
All the cases we considered in Section~4 belong to this class of metric,
except for the sphere and pseudo-sphere.
Thus, the supersymmetry has to be responsible for removing the high degeneracy
of the eigenstates presented in the ordinary Hall effect.

A second
unpleasant feature of our system is the absence of a confining potential.
To accommodate such a potential, auxiliary bosonic degrees of freedom are
needed, which requires adding extra supermultiplets to our present scheme.
Such multiplets will bring in new physical fermions, and we have not yet
a recipe how to deal with those.

Geometrically, the absence of SO(5) invariance in our systems originates
from treating the four-dimensional coordinates $y_\mu$ as SO$(5)/$SO(4)
coset-space coordinates, and so a part of the SO(5) symmetry is non-linearly
realized via
\be\label{soo5}
\delta y_\mu\=\frac{ \rho-y^2}{2}\,a_\mu + \left(a_\nu y^\nu\right)\,y_\mu\ .
\ee
While the four-sphere possesses such an invariance, the constraints defining
the ${\cal N}{=}4$ hypermultiplet do not. Instead, as directly follows
from~\cite{IKL1}, the physical bosons of the hypermultiplet parametrize
$\mathbb{R}\times \mathbb{S}^3$, a space which cannot carry SO(5)~invariance.
A possible solution could be to replace the hypermultiplet by some nonlinear
supermultiplet, whose bosonic components should parametrize the four-sphere.
The most natural candidate for this role is the nonlinear $(4,8,4)$
supermultiplet~\cite{nlm,IVAN}, which would extend the number of fermions
to eight.
Nevertheless, we do not expect the corresponding action to enjoy ${\cal N}{=}8$
supersymmetry~\cite{Braz}, due to rather strong restrictions on the bosonic
metric imposed by the four extra supersymmetries. We are planning to consider
these possibilities in more detail elsewhere.

\section*{Acknowledgements}
We are grateful to A.~Nersessian for his collaboration during part of this work.
We also thank D.~Sorokin for fruitful discussions.
S.K. thanks the ITP at Leibniz Universit\"{a}t Hannover for hospitality during
finalizing of this work. A.S. thanks Padova University for hospitality.
This work was partially supported by the grants
09-02-01209, 09-02-91349, and Volkswagen Foundation
grant I/84 496.

\end{document}